\documentclass[11pt]{article}
\usepackage{amssymb}
\usepackage{epsfig}
\def \beq{\begin{equation}}
\def \eeq{\end{equation}}
\def \beqa{\begin{eqnarray}}
\def \eeqa{\end{eqnarray}}
\def \beqal{\begin{eqletters}\begin{eqnarray}}
\def \eeqal{\end{eqnarray}\end{eqletters}}

\def\sg{\hat g}
\def\lamB{{\lambda}}

\def\DY{\Delta Y}

\begin{document}
\author{ \bf Wojciech S\l omi\'nski 
\\
Institute of Physics, Jagellonian University,
\\ 
Reymonta 4, 30-059 Krak\'ow, Poland
\\ and
\\
\bf Jerzy Szwed \footnote {Work supported by
the European Commission contract ICA1-CT-2002-70013/INCO Strategic action on training and excellence, 5FP}
\\
Institute of Physics, Jagellonian University,
\\ 
Reymonta 4, 30-059 Krak\'ow, Poland
\\and
\\
Centre de Physique Th\'eorique, CNRS Luminy Case 907, \\
13288 Marseille Cedex 09, France 
}
\title{\small CENTRE DE PHYSIQUE TH\'EORIQUE \footnote{Unit\'e Mixte de Recherche du CNRS et des Universit\'es de Provence, de la M\'editerran\'ee et du Sud Toulon-Var - Laboratoire affili\'e a la FRUNAM - FR 2291}
\\
CNRS-Luminy, Case 907
\\
13288 Marseille Cedex 9
\\
FRANCE
\vskip2cm
\Large \bf Spin Effects in the Electron Structure Functions}
\maketitle

TPJU - 8/2004

CPT-2004/P.032
\begin{abstract}
The electron structure functions are studied in polarized $e^+e^-$ scattering. The formulae for longitudinally and transversely polarized electrons are presented. The smallnes of the electron mass leads to negligible cross-sections and asymmetries in some cases.
Positivity constraints on the structure functions and 
parton densities are constructed and discussed. The cross-section asymmetries at very high energies, where the inclusion of all elecroweak bosons is necessary, 
are calculated. Numerical examples, using the asymptotic solutions for the parton densities inside the electron, are presented. 

\end{abstract}

\newpage
\section{Introduction.}

The electron structure function has been constructed and studied in a series of papers
 \cite{SSPL}-\cite{EPJ} as well as first measured at LEP \cite{Muryn}. It turned to be an useful notion at present energies where 
it can serve, among others, as a cross-check of the extensively studied photon structure function \cite{resph, resphexp}. 
At very high energies its use is without doubt necessary: it includes non-negligible contributions from 
all intermediate bosons and their  ($\gamma$-$Z$) interference. A detailed comparison with the "photon structure 
approach" has been presented in Ref. \cite{EPJ}.
In this paper we study the spin dependence of the partonic content of the electron/positron in 
polarized deep-inelastic $e^+e^- \rightarrow eX$ scattering. The structure of the process can be visualized 
as in Fig. 1. The electron $e_\lambda$ of polarization $\lambda =\pm $ and momentum 
$l=(E,\vec l)$ probes the positron $e_\kappa$ of polarization $\kappa =\pm $ and momentum 
$k$  by emitting the photon or $Z$ boson (The reverse setup where the positron probes the electron is analogical but differs quantitatively when weak interactions are included.). The scattered electron is tagged 
so that we can determine the relevant variables by measuring  its final momentum 
$l'=(E_{\rm tag},\vec l')$ and scattering angle $\theta _{\rm tag}$. The target positron remains untagged - its 
momentum transfer $P^2 = - {(k-k')}^2$ is smaller than experimentally determined upper limit $P^2_{max}$. The hadronic structure developped by the target lepton begins with the emmission of colinear electroweak boson: photon, $Z$ or $W$ (The anti-tag condition cannot be imposed when the $W$ boson is produced.).

In Section 2. we limit ourselves to the energy range where both the probe and target leptons emit photons only. We introduce there the cross-sections and asymmetries in analogy to the electron-nucleon scattering. The positivity constraints among structure functions and parton densities inside the electron are presented in Section 3. In Section 4 we generalize the discussion to the case where  all electroweak bosons are taken into account. We include the $\gamma$, $Z$ and $\gamma - Z$ emission by the probe lepton and the contribution of the $\gamma$, $Z$, $\gamma - Z$ and $W$ to the lepton structure function. We illustrate the calculation with numerical examples where we take the parton densities inside the electron from our asymptotic solutions  of the evolution equations \cite{SSPL}-\cite{EPJ}. Section 5 summarizes our results.

\begin{figure}
\centerline{\epsfig{file=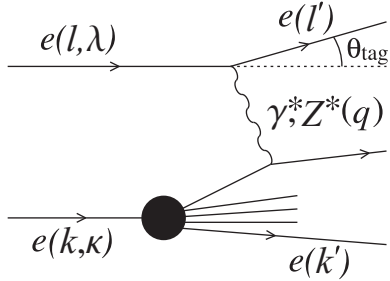,width=2in}}%
\caption{The electron-positron scattering and the electron structure}
\label{fig1}
\end{figure}

\vskip1cm
\section{Cross-sections and asymmetries.}

Defining the deep inelastic scattering variables in the standard way:
\beqa
 Q^2=-(l-l')^2=2\,E\,E_{\rm tag}(1-\cos \theta_{\rm tag}),
\\
\nonumber
z={Q^2\over {2\,k\,q}}=
{\sin^2(\theta_{\rm tag}/2) \over E/E_{\rm tag} -\cos^2(\theta_{\rm tag}/2) }
\label{QQ}
\eeqa
the cross-section for unpolarized electron-positron scattering can be written in the limit $Q^2 >> P_{max}^2$:
\beqa
\label{ele}
{{d^2\sigma _{ee}}\over {dzdQ^2}}
  \equiv 
 {1\over 2} \,{{{d^2\sigma _{e_+ e_+}}  + 
{d^2\sigma _{e_- e_+}}}\over {dzdQ^2}}
\nonumber
\eeqa
\beq
 = 
 {2\pi \alpha^2\over z Q^4}
\Big[
(1  +  (1-y)^2) \, F^e_2(z,Q^2,P^2_{\rm max})
 -y^2 \, F_L^e(z,Q^2,P^2_{\rm max})
\Big].
\eeq
with 
\beqa
y =
{{qk}\over {lk}}
=
 1-(E_{\rm tag}/E)\cos ^2(\theta_{\rm tag}/2).
\eeqa
Taking into account the smallness of the electron mass $m$, the longitudinal structure function reads:
\beqa
F_L(z,Q^2,P^2_{\rm max}) &=& {1 \over {2z}}({{4m^2z^2} \over {Q^2}} +1) F_2(z,Q^2,P^2_{\rm max}) - F_1(z,Q^2,P^2_{\rm max})
\nonumber \\
&\simeq &
{1 \over {2z}} \, F_2(z,Q^2,P^2_{\rm max}) - F_1(z,Q^2,P^2_{\rm max})
 \nonumber \\
 &\equiv&
 R \,  F_1(z,Q^2,P^2_{\rm max}).
\eeqa
and the cross-section Eq. (\ref{ele}) reduces to
\begin{eqnarray}
\label{unpol}
{{d^2\sigma _{ee}}\over {dzdQ^2}} &=&
 {4\pi \alpha^2\over  Q^4}
\bigg[
\Big(1+(1-y)^2\Big)(1+R)-{1\over {2z}}y^2R \bigg] \,  F^e_1(z,Q^2,P^2_{\rm max}).
\end{eqnarray}
The structure function $F_1$ is related to the quark densities $q_i$ inside the electron in the standard way:
\beqa
\label{F1eg}
F_1(z,Q^2,P^2_{\rm max})= \sum_i e^2_{q_i} \,  q_i(z,Q^2,P^2_{\rm max})
\,,
\eeqa
where $e_{q_i}$ is the $i$-th quark fractional charge.
The polarized cross-section reads

\beqa
\label{elehel}
{{d^2\Delta \sigma _{ee}}\over {dzdQ^2}}
&\equiv&
 {1\over 2} \,  {{{d^2\sigma _{e_+ e_+}}- 
{d^2\sigma _{e_- e_+}}}\over {dzdQ^2}}
\nonumber 
\\
&=& 
 {-4\pi \alpha^2\over  Q^4}
\Big[
(1-(1-y)^2) \, g_1(z,Q^2,P^2_{\rm max})
 -2 \, z \, y \, \epsilon \,  g_2(z,Q^2,P^2_{\rm max})\Big]
\nonumber 
\\
& \simeq &
{-4\pi \alpha^2\over  Q^4}
\Big(1-(1-y)^2\Big) \, g_1(z,Q^2,P^2_{\rm max})
\eeqa
where we introduced 
\beq
\epsilon \,  \equiv  \, {{2m^2}\over s} \,  {\buildrel \rm LAB \over =} \,  {m \over E_{\rm LAB}}  \, {\buildrel \rm CM \over =} \,  {m^2 \over {2E_{\rm CM}}} \, 
\simeq 0
\eeq
to demosntrate the terms present in the electron - nucleon scattering and negligible in the present case due to the smallness of the electron mass.

Polarizing  the spin $\vec S$ of the probing electron in the transverse direction one arrives at negligible cross-section ($\sim {\sqrt{\epsilon}})$, again due to the smallness of the electron mass 
(the angle $\phi$ is the angle beetween the $(\vec k,\vec k')$ 
and $(\vec k,\vec S )$ planes):
\beqa
\label{eletrans}
{{d^2\Delta_{\perp} \sigma _{ee}}\over {dzdQ^2}}
& \equiv & 
 {1\over 2} {{{d^2\sigma _{e\uparrow e_+}}- 
{d^2\sigma _{e\downarrow e_+}}}\over {dzdQ^2}}
\\
\nonumber
= 
 {-4\pi \alpha^2\over  Q^4}
\sqrt{2zy(1-y)}  &\sqrt{\epsilon} & \Big(y \, g_1(z,Q^2,P^2_{\rm max})+2 \, g_2(z,Q^2,P^2_{\rm max}) \Big)\cos {\phi} 
\eeqa

Taking into account the avove results, the discussion of relevant asymmetries simplifies. The spin-spin asymmetries used in deep-inelastic scattering:
\beqa
\label{ALL}
A_{\parallel}
=
\bigg({{{d^2\Delta \sigma _{ee}}\over {dzdQ^2}}\bigg) \bigg/
\bigg({{d^2\sigma _{ee}}\over {dzdQ^2}}}\bigg)
\quad \mathrm{and} \quad
A_{\perp}
=
\bigg({{{d^2\Delta_{\perp} \sigma _{ee}}\over {dzdQ^2}}\bigg)\bigg/
\bigg({{d^2\sigma _{ee}}\over {dzdQ^2}}}\bigg)
\eeqa
can be  expressed in terms of virtual photon scattering asymmetries $A_{1,2}$ (to compare with the nucleon case see eg. Ref. \cite{leader}) :
\beqa
A_{\parallel} = D(A_1 + \eta A_2) 
\quad {\rm and} \quad
A_{\perp} = d (A_2 + \xi A_1)
\eeqa
where 
\beqa
A_1 = {{g_1 - {{2z}\over y} \, \epsilon \,  g_2}\over F_1} \simeq {g_1 \over F_1}
&,& \quad
A_2 = {\sqrt{{2z}\over y}  \, \sqrt{\epsilon}  \, {{g_1+g_2}\over F_1}} \simeq 0 ,
\nonumber
\\
D = {{1-\kappa (1-y)}\over {1+\kappa R}}
&,& \quad d = D\sqrt{2\kappa\over {1-\kappa}} ,
\nonumber
\\
\kappa = {{1+(1-y)^2}\over {2(1-y)}} &,& \quad 
\xi =\eta {{1+\kappa}\over {2\kappa}} \simeq 0 ,
\eeqa
\beq
\nonumber
\eta = {{\sqrt{2zy} [1+(1-y)^2]}\over {(1-y)[1-(1-y)^2]}} \,  \sqrt{\epsilon} \simeq 0 .
\eeq
From the above definitions one sees that the structure function $g_1$ is directly measuerd by the asymmetry $A_{\parallel}$ whereas the structure function $g_2$ is practically inaccesible in $e^+e^-$ scattering.
 
\vskip1cm

\section{Positivity constraints. } 

The structure functions $F_{1,2}$ and $g_{1,2}$ are related to the forward Compton scattering amplitudes $M_{a\lambda b\mu }$ and photoabsprbtion cross-sections $\sigma^{(\pm)}_{T,S,ST}$. In our notation $a(b)$ and $\lambda (\mu )$ are the helicities of the initial (final) photon and positron:
\beqa
\label{photoabs}
{1\over m}F_1 &=& {1\over 2}(M_{1+1+} +M_{1-1-}) \sim \sigma^+_T
\nonumber
\\ 
{1\over m}F_L &=& M_{0+0+} \sim \sigma_S
\nonumber
\\ 
{1\over m}{g_1 - {{2z}\over y}\epsilon g_2} &=& {1\over 2}(M_{1+1+}-M_{1-1-}) \sim \sigma^-_T
\nonumber
\\ 
{2\sqrt{z}\over my}  \, \sqrt{\epsilon} \,  (g_1+g_2) &=& M_{1-0+} \sim \sigma_{TS}
\eeqa

The structure of the above relations is analogical to that of the deep-inelastic electron nucleon scattering due to the identical spin structure of both processes. Therefore we can easily derive the constraints on the structure functions coming from the positivity of the photoabsorbtion cross-sections. The first group can be written as:
\beqa
\label{ineq1}
F_1  &\geq &  0
\nonumber
\\
F_L  =  {1 \over {2z}}({{4m^2z^2} \over {Q^2}} +1) \,  F_2 - F_1 & \geq & 0,
\nonumber
\\
\sqrt{{R(1+A_1)\over 2}}  & \geq  & A_2 .
\eeqa
The second type of relation introduces constraints among the parton densities inside the electron and the chirality-odd, twist 2 structure function $h_1$ \cite{ralston}.  Derived first by Soffer \cite{soffineq} for the nucleon, the relation takes analogical form in the electron case:
\beq
\label{ineq2}
q(z)+\Delta q(z) \,  \geq  \, 2 \mid h_1(z) \mid .
\eeq
The inequality ({\ref{ineq2}) holds for each quark (antiquark) flavour separately.
It seems to be fulfilled trivially since $h_1$, representing the parton transversity distributions, does not evolve from the initial electrons trough the photons.

\section{Cross-section and asymmetries with all electroweak bosons.}

In the presence of all electroweak bosons the picture of $e^+e^-$ scattering becomes more complicated than in the pure QED approach. The structure functions $F_1$ and $g_1$ depend in general on both the polarization of the probe and the target. In addition, the deep inelastic  scattering depends now on which lepton is the target and which is the probe. But as in the previous sections, due to the smallness of the electron mass, many terms can be 
omitted without loss of accuracy. Out of all possible polarization directions only that along the momentum of the leptons is nonnegligible. 

The cross-section for the scattering of electron of helicity $\lambda={\pm}$ off the 
positron of helicity $\kappa={\pm}$ reads:
\beqa
\label{lamkap}
{d\sigma(e_\lamB e_\kappa) \over dz\,dQ^2} 
=
  4\pi \alpha^2
\sum_{B,B'}
 {\sg_{Be_\lamB} \over Q^2 + M_B^2} {\sg_{B'e_\lamB} \over Q^2 + M_{B'}^2}
  \left[
     Y_{+}(y) J_{BB'}^{\lamB \kappa}
    + Y_{-}(y) J_{BB'}^{-\lamB \kappa}
  \right]
\eeqa
where
$M_B$, $M_B'$ are the masses of exchanged bosons ($\gamma$ or $Z$), 
$\sg_{Be_\lamB}$ - the  lepton-boson electroweak couplings,  
\beqa
 Y_-(y) = 1
\quad , \quad
 Y_+(y) = (1-y)^2
\eeqa
and
\begin{eqnarray}
 J_{BB'}^{\lambda \kappa}
=
 \sg_{Bq_{-\lambda}} \sg_{B'q_{-\lambda}} f^{e_\kappa}_{q_{-\lambda}}(z,Q^2,P^2_{max})
+
 \sg_{Bq_\lambda} \sg_{B'q_\lambda} f^{e_\kappa}_{\bar q_{-\lambda}}(z,Q^2,P^2_{max}) . 
\end{eqnarray}
In the last formula  $f^{e_\kappa}_{q_{\lambda}}(z,Q^2,P^2)$ represent the densities of quarks of helicity $\lambda$ inside the positron of helicity $\kappa$.

By  defining 
\beqa
Y = Y_+ +  Y_- \quad ,
\quad
\DY = Y_+ - Y_-
\eeqa

and 
\beqa
  F^{\lamB \kappa}_1 &=&
  \sum_{B,B'}
  \sg_{Be_\lamB} \sg_{B'e_\lamB}
  {Q^2 \over Q^2 + M_B^2} { Q^2 \over Q^2 + M_{B'}^2}
  \left[
  J_{BB'}^{+ \kappa} + J_{BB'}^{- \kappa}
  \right] ,
\\
  g^{\lamB \kappa}_1 &= \lamB &
  \sum_{B,B'}
  \sg_{Be_\lamB} \sg_{B'e_\lamB}
  {Q^2 \over Q^2 + M_B^2} { Q^2 \over Q^2 + M_{B'}^2}
  \left[
    J_{BB'}^{+ \kappa} - J_{BB'}^{- \kappa}
  \right] 
\eeqa
we can write the cross-section (\ref{lamkap}) as follows:
\beqa
\hspace{-8mm}
  {d\sigma(e_\lamB e_\kappa) \over dz\,dQ^2} 
  &\equiv&
  {2\pi \alpha^2 \over Q^4}
  \left[
    Y(y) F^{\lamB \kappa}_1 + \DY(y) g_1^{\lamB \kappa}
  \right]
\eeqa

In pure QED case, where the only boson contributing to the process is the photon (both the highly 
virtual from the beam electron
and the nearly real from the target positron), the cross-section reduces to the formulae (\ref{ele}) and (\ref{elehel}).

To see what effects introduces the inclusion of all electroweak bosons we calculated the above cross-sections numerically using as the parton densities - the asymptotic solutions of our evolution equations \cite{SSAPP}-\cite{EPJ}. In Fig. 2 the $e^-e^+$ scattering cross-section is presented at $\sqrt{s} = 1 TeV, Q^2 = 5\times10^5 GeV^2 and  P^2_{max}=2\times10^3 GeV^2$ for various helicity configurations. Large differences result from different electroweak couplings of $\gamma, W$ and $Z$ bosons. In Fig 3. we present the largest asymmetry which can be constructed from the above cross-sections 
\beq
\label{asymmetryA}
A \,=\, \bigg({{{d\sigma(e_+ e_-)  - d\sigma(e_- e_+)} \over dz\,dQ^2} \bigg) \bigg/
  \bigg( {{{d\sigma(e_+ e_-)  + d\sigma(e_- e_+)} \over dz\,dQ^2}}} \bigg) .
\eeq

 For comparison we present the same asymmetry with the photon only. The presence of weak bosons is clearly seen.

\begin{figure}
\centerline{\epsfig{file=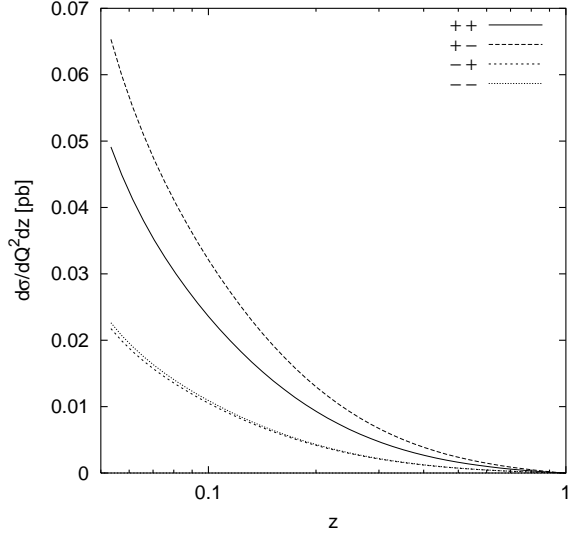,width=3in}}%
\caption{The cross-section ${d\sigma(e_\lamB e_\kappa) \over dz\,dQ^2}$ as a function of $z$ for $\sqrt{s} = 1 TeV, Q^2 = 5\times10^5 GeV^2, P^2_{max}=2\times10^3 GeV^2$ with the (electron, positron) helicity $(\lambda, \kappa)$. }
\label{fig1}
\end{figure}

\begin{figure}
\centerline{\epsfig{file=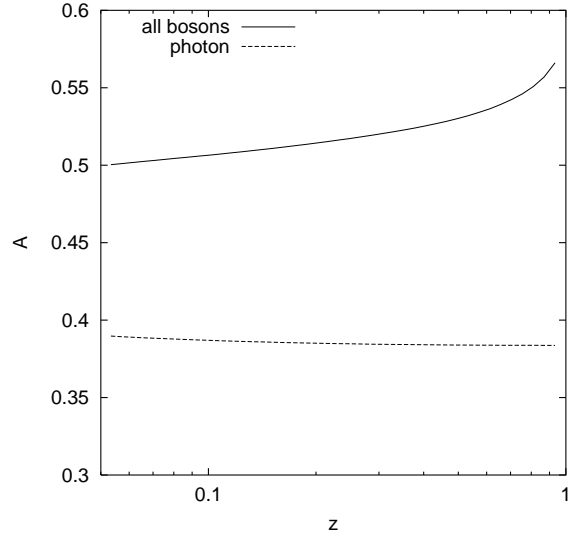,width=3in}}%
\caption{The asymmetry Eq.(\ref{asymmetryA}) as a function of z with all electroweak bosons included in the electron structure function (solid line) and the photon only (dashed line).}
\label{fig1}
\end{figure}

\section{Summary}

We have completed the analysis of the QCD structure of leptons by looking at the spin effects which appear in $e^+e^-$ collisions at present and $TeV$ energies. As compared to the electron-nucleon scattering, where the spin structure of the process is analogical, many formulae simplify due to the smallness of the electron mass. On the other hand new effects can be observed at very high energies, in particular  all electroweak gauge bosons play an important role in building up the QCD structure. The use of the electron/positron structure functions which  include also  interference effects is then necessary. We also give numerical estimates what is the size of these new effects at $TeV$ energies. Since no parton parametrisations exist at these energies, we use our asymptotic solutions of the corresponding evolution equations. The contribution from weak gauge bosons turns out to be quantitatively significant and should be visible in the next generation of $e^+e^-$ experiments.

\vskip1cm

\section{Acknowledgements.}

J.S.  would like to thank Jacques Soffer for discussions and  
Centre de Physique Th\'eorique in Marseille for hospitality.


\end{document}